\documentclass[12pt]{article}
\usepackage{amsfonts}
\usepackage{amssymb}
\usepackage{amsbsy}
\usepackage{color}

\input epsf.sty

\newlength{\TZ}
\newcommand{\BEQ}{\begin{equation}}     
\newcommand{\BEA}{\begin{eqnarray}}
\newcommand{\BD}{\begin{displaymath}}
\newcommand{\EEQ}{\end{equation}}       
\newcommand{\EEA}{\end{eqnarray}}
\newcommand{\ED}{\end{displaymath}}
\newcommand{\vph}{\varphi}              
\newcommand{\D}{{\rm d}}                
\newcommand{\demi}{\frac{1}{2}}         
\newcommand{\wit}[1]{\widetilde{#1}}    

\renewcommand{\vec}[1]{\boldsymbol{#1}} 

\newcommand{\R}{\mathbb{R}}

\newcommand{\Z}{\mathbb{Z}}

\catcode`\@=11
\def\numberbysection{\@addtoreset{equation}{section}
        \def\theequation{\thesection.\arabic{equation}}}
\numberbysection

\begin{document}

\begin{titlepage}

\vskip 1.5 cm
\begin{center}
{\bf \LARGE From conformal invariance towards \\[0.17truecm] dynamical symmetries of the collisionless Boltzmann equation}
\end{center}

\vskip 2.0 cm
\centerline{{\bf Stoimen Stoimenov}$^a$ and {\bf Malte Henkel}$^b$}
\vskip 0.5 cm
\centerline{$^a$ Institute of Nuclear Research and Nuclear Energy, Bulgarian Academy of Sciences,}
\centerline{72 Tsarigradsko chaussee, Blvd., BG -- 1784 Sofia, Bulgaria}
\vspace{0.5cm}
\centerline{$^b$ Groupe de Physique Statistique, D\'epartement de Physique de la Mati\`ere et des Mat\'eriaux,}
\centerline{Institut Jean Lamour (CNRS UMR 7198), Universit\'e de Lorraine Nancy,}
\centerline{B.P. 70239, F -- 54506 Vand{\oe}uvre l\`es Nancy Cedex, France}

\begin{abstract}
Dynamical symmetries of the collisionless Boltzmann transport equation, or Vlasov equation,
but under the influence of an external driving force, are derived from
non-standard representations of the $2D$ conformal algebra. In the case without external forces, the symmetry of the
conformally invariant transport equation is
first generalised by considering the particle momentum as an independent variables. This new conformal
representation can be further extended to include an external force.
The construction and possible physical applications are outlined.
\end{abstract}
\end{titlepage}

\setcounter{footnote}{0}

\section{ Introduction}

The {\em Boltzmann transport equation} (BTE) \cite{Boltzmann,Kreuz81,Haug97,4}
furnishes a semi-classical description of the effects of particle transport, including the influence of external forces,
on the effective single-particle 
distribution function $f=f(t,\vec{r},\vec{p})$ of a small cell in phase phase, centred at position $\vec{r}$ and
momentum $\vec{p}$. For a system with identical particles of mass $m$, the Boltzmann equation reads
\BEQ
\frac{\partial f}{\partial t}+\frac{\vec{p}}{m}\cdot\frac{\partial f}{\partial \vec{r}}+\vec{F}\cdot\frac{\partial f}{\partial \vec{p}}
=\left(\frac{\partial f}{\partial t}\right)_{\rm coll}.
\label{inboltzmann}
\EEQ
Here, $\D N = f(t,\vec{r}, \vec{p},)\D\vec{r}\,\D\vec{p}$
is the number of particles in a cell of phase volume $\D\vec{r}\,\D\vec{p}$, centred at position $\vec{r}$ and momentum
$\vec{p}$ \cite{4}. In addition, $\vec{F}=\vec{F}(t,\vec{r})$ is the force field acting on the particles in the fluid.
The term on the right-hand-side is added to describe the effect of collisions between particles. It is a statistical term and requires knowledge
of the statistics the particles obey, like the Maxwell-Boltzmann, Fermi-Dirac or Bose-Einstein distributions.
In his famous `Sto{\ss}zahlansatz' (or hypothesis of molecular chaos), Boltzmann obtained an explicit form for it.
In a modern notation, for example for an interacting Fermi gas, where a particle from a state with momentum $\vec{p}$ is scattered to a state with momentum $\vec{p}'$, whereas a second particle is scattered from
a momentum $\vec{q}$ to a momentum $\vec{q}'$, the collision term reads
\BEA
\left(\frac{\partial f}{\partial t}\right)_{\rm coll} &=&
-\int\!\D\vec{p}'\D\vec{q}\D\vec{q}'\: w(\{\vec{p},\vec{q}\}\to\{\vec{p}',\vec{q}'\})
\nonumber \\
& &\times \left[ f(\vec{p}) f(\vec{q}) (1-f(\vec{p}'))(1-f(\vec{q}')) - f(\vec{p}') f(\vec{q}') (1-f(\vec{p}))(1-f(\vec{q})) \right]
\nonumber
\EEA
where $w(\{\vec{p}\vec{q}\}\to\{\vec{p}'\vec{q}'\})$ is the normalised transition probability from the two-particle state with
momenta $\{\vec{p},\vec{q}\}$ to the state labelled by $\{\vec{p}',\vec{q}'\}$. Clearly, solving this widely studied equation is a very difficult task.
It might be hoped that symmetries could be helpful. The equation without the collision term is known as the {\em Vlasov equation} \cite{Vlasov38}.
Relationship with Landau damping and a physicists' derivation can be found in \cite{Vilani14,Elskens14}.
In this work, we shall explore  a class of symmetries of the (collisionless) BTE.

Throughout, we shall restrict to $d=1$ space dimension\footnote{By analogy with other constructions of local scale symmetries, 
see \cite{Henkel02,Duval09,Martelli10,Henkel14} and especially \cite{Henkel10} and refs. therein, 
we expect a straightforward extension of the results reported here to $d>1$. 
Since we shall construct here a finite-dimensional Lie algebra of dynamical conformal symmetries of the $1D$ collisionless BTE, 
one should indeed expect that an extension to $d>1$ exists. That symmetry algebra should
contain three generators $X_{\pm 1,0}$, along with a vector of generators $\vec{Y}_n$ and also spatial rotations.}.
We start from a non-standard representation, isomorphic to the infinite-dimensional Lie algebra of conformal transformations 
in $d=2$ dimensions.\footnote{For the sake of clarify, we shall adopt the following convention of terminology: the infinite-dimensional Lie algebra 
$\left\langle X_n, Y_n\right\rangle_{n\in\mathbb{Z}}$ will be called a (centreless) `{\em conformal Virasoro algebra}'. Its maximal finite-dimensional
sub-algebra $\left\langle X_n, Y_n\right\rangle_{n\in\{-1,0,1\}}$ will be called a `{\em conformal algebra}'.} This Lie algebra is spanned by the generators $\langle X_n, Y_n\rangle_{n\in \Z}$ and can be defined from the commutators
\cite{Henkel02,Henkel10}
\BEQ
[X_n,X_{m}] = (n-m)X_{n+m},\quad  [X_n,Y_{m}] = (n-m)Y_{n+m},\quad [Y_n,Y_{m}] = \mu (n-m)Y_{n+m}
\label{commutators}
\EEQ
where $\mu$ is a parameter. An explicit realisation in terms of time-space transformation is \cite{Henkel02,Henkel10}:
\BEA
X_n   &=& -t^{n+1}\partial_t-\mu^{-1}[(t+\mu r)^{n+1}-t^{n+1}]\partial_r-(n+1)xt^n- (n+1)\frac{\gamma}{\mu}[(t+\mu r)^{n}-t^{n}]\nonumber\\
Y_{n} &=& -(t+\mu r)^{n+1}\partial_r- (n+1)\gamma (t+\mu r)^{n}
\label{infinivarconf}
\EEA
such that $\mu^{-1}$ can be interpreted as a velocity (`speed of light/sound') and where $x,\gamma$ are constants.\footnote{The contraction
$\mu \to 0$ of (\ref{infinivarconf}) produces the non-semi-simple
`altern-Virasoro algebra' $\mathfrak{altv}(1)$ (but without central charges). Its maximal finite-dimensional sub-algebra is
the conformal galilean algebra $\mathfrak{alt}(1)\equiv\mbox{\sc cga}(1)$ \cite{Henkel02,Henkel03a}, see also \cite{Duval09,Martelli10}.
The \mbox{\sc cga}(d) is non-isomorphic to either the standard Galilei algebra or else the Schr\"odinger algebra.}
Writing $X_n=\ell_n +\bar{\ell}_n$ and $Y_n=\mu^{-1}\bar{\ell}_n$, where the generators $\langle \ell_n, \bar{\ell}_n\rangle_{n\in\Z}$ satisfy
$[\ell_{n},\ell_m]=(n-m)\ell_{n+m}, [\bar{\ell}_n,\bar{\ell}_m]=(n-m)\bar{\ell}_{n+m}, [\ell_n,\bar{\ell}_m]=0$,
it can be seen that, provided $\mu\ne 0$, the above Lie algebra (\ref{commutators})
is isomorphic to a pair of Virasoro algebras
$\mathfrak{vect}(S^1)\oplus\mathfrak{vect}(S^1)$ with a vanishing central charge. However, this isomorphism does not imply that physical
systems described by two different representations of the conformal Virasoro algebra, or the conformal algebra, with commutators (\ref{commutators}), 
were trivially related.
For example, it is well-known that if one uses the generators of the standard representation of conformal invariance or else the non-standard
representation (\ref{finitvarconf}) in order to find co-variant two-point functions, the resulting scaling forms are different \cite{Henkel02}.

Now, consider the maximal finite-dimensional sub-algebra $\langle X_{\pm 1,0}, Y_{\pm 1,0}\rangle$, which for $\mu\ne 0$ in turn is isomorphic to the
direct sum $\mathfrak{sl}(2,\R)\oplus\mathfrak{sl}(2,\R)$. The explicit realisation follows from from (\ref{infinivarconf})
\BEA
X_{-1} &=& -\partial_t, \quad X_0 = -t\partial_t-r\partial_r-x, \quad
~X_{1} = -t^2\partial_t-2tr\partial_r-\mu r^2\partial_r-2xt-2\gamma r\nonumber\\
Y_{-1} &=& -\partial_r, \quad Y_0=-t\partial_r-\mu r\partial_r-\gamma, \quad
Y_1 = -t^2\partial_r-2\mu tr\partial_r-\mu^2r^2\partial_r-2\gamma t - 2\gamma\mu r~~~~
\label{finitvarconf}
\EEA
Here, the generators $X_{-1}, Y_{-1}$ describe time- and space-translations, $Y_{0}$ is a (conformal) Galilei transformation,\footnote{Since the
commutator $[Y_{0}, Y_{-1}]$ does not vanish and does not give a central element of the Lie algebra (\ref{commutators}),
its structure is fundamentally different from algebras containing the usual Galilei algebra as a sub-algebra.} $X_{0}$ gives the
dynamical scaling $t\mapsto \lambda t$ of $r\mapsto \lambda r$  (with $\lambda\in\R$) such that the so-called `dynamical exponent' $z=1$
since both time and space are re-scaled in the same way and finally $X_{+1}, Y_{+1}$ give `special' conformal transformations.
In the context of statistical mechanics of conformally invariant phase transitions,
one characterises co-variant quasi-primary scaling operators through the invariant parameters $(x,\mu,\gamma)$,
where $x$ is the scaling dimension.

Finally, the finite-dimensional representation (\ref{finitvarconf}) acts as a dynamical symmetry on the equation of motion
\BEQ
{\hat S}\phi(t,r)=(-\mu\partial_t+\partial_r)\phi(t,r)=0. \label{ineq1}
\EEQ
in the sense that a solution $\phi$ of ${\hat S}\phi=0$ is mapped onto another solution of the same equation. Indeed, it is easily checked that
$[{\hat S},Y_{\pm 1,0}] = [{\hat S}, X_{-1}]=0$ and
\BEQ \label{dynsym}
{} [{\hat S},X_0] = -{\hat S},\quad [{\hat S},X_{1}] = -2t{\hat S}+2(\mu x-\gamma)
\EEQ
It follows that for fields $\phi$ with scaling dimensions $x_{\phi}=x=\gamma/\mu$ the algebra (\ref{finitvarconf}) really leaves the solution space of
the equation (\ref{ineq1}) invariant.\footnote{Since $[{\hat S},X_n]=-(n+1)t^n{\hat S} +n(n+1)t^{n-1}(\mu x-\gamma)$ and $[{\hat S},Y_n]=0$, for all
$n\in\mathbb{Z}$, this symmetry extends to the centreless Virasoro algebra (\ref{commutators}).}

In order to return to the Boltzmann equation, we consider eq.~(\ref{ineq1}) in the form
\BEQ
{\hat L}f=(\mu\partial_t+v\partial_r)f(t,r,v)=0 \label{ineq2}
\EEQ
where $f=f(t,r,v)$ is interpreted as a 
single-particle distribution function and where we consider $v$ as an additional variable. Eq.~(\ref{ineq2}) is a
simple Boltzmann (or Vlasov) equation, without an external force and without a collision term, and in one space dimension. From (\ref{dynsym}),
with $v$ fixed (and normalised to $v=1$), 
its solution space is conformally invariant\footnote{With respect to  eq.~(\ref{ineq1}), $\mu\mapsto -\mu$ was replaced.
This change must also be made in the generators (\ref{finitvarconf}) and commutators (\ref{commutators}).}. In section~2, we shall generalise
the above representation of the conformal algebra to the situation with $v$ as a further variable. In section~3, we shall further extend this
to the case when an external force $F=F(t,r,v)$, possibly depending on time, spatial position and velocity, is included. The aim of these calculations
is to determine which situations of potential physical interest with a non-trivial conformal symmetry might be identified.
This explorative study aims at identifying lines for further study, 
which might lead later to a more comprehensive understanding of the possible  symmetries of
Boltzmann equations. Taking into account the collision term is left for future work. We shall concentrate on $d=1$ space dimension throughout.
Conclusions and final comments are given in section~4.

\section{Collisionless Boltzmann equation without external forces}

In our construction of conformal dynamical symmetries of the $1D$ collisionless BTE, we shall often meet Lie algebras of a certain structure.
These will be isomorphic to the two-dimensional conformal algebra.

\noindent
{\bf Proposition 1:} {\it The Lie algebra $\langle X_{n}, Y_{n}\rangle_{n\in\Z}$ defined by the commutators}
\BEQ 
{}[X_n, X_m] = (n-m) X_{n+m},\quad [X_n, Y_m] = (n-m) Y_{n+m},\quad [Y_n, Y_m] = (n-m) \left( k X_{n+m} + q Y_{n+m} \right)
\label{symmetryalgebra}
\EEQ
{\it where $k,q$ are constants, is isomorphic to the pair of centreless Viraso algebras $\mathfrak{vect}(S^1)\oplus\mathfrak{vect}(S^1)$.}\\

\noindent {\bf Proof:} 
For either $k=0$ or $q=0$ this is either evident or else has already been seen in section~1. In  the other case, consider the change of basis
$X_n=\ell_n +\bar{\ell}_n$ and $Y_n=\alpha\ell_n - \beta\bar{\ell}_n$ where $\ell_n,\bar{\ell}_n$ are two families of commuting generators of
$\mathfrak{vect}(S^1)$ and $\alpha$ and $\beta$ are constants such that $\alpha+\beta\ne0$. It then follows $k=\alpha\beta$ and $q=\alpha-\beta$.
\hfill ~ q.e.d. 

This implies in particular the isomorphism of the maximal finite-dimensional sub-algebras, or `conformal algebras' in the terminology chosen here. 
By definition, this `conformal algebra' obeys the commutators (\ref{symmetryalgebra}), but with $n,m\in\{-1,0,1\}$. 

Our construction of dynamical symmetries of the equation (\ref{ineq2}) 
follows the lines of the construction of local scale-invariance in time-dependent
critical phenomena \cite{Henkel02}. The physically motivated requirements are:
First of all it is clear that the equation is invariant under time-translations:
\BEQ
X_{-1}=-\partial_t, \quad [{\hat L},X_{-1}]=0 \label{timetranslations}
\EEQ
Some kind of dynamical scaling must be present as well. Its most general form is
\BEQ
X_0=-t\partial_t-\frac{r}{z}\partial_r-\frac{1-z}{z}v\partial_v-x, \quad [{\hat L},X_0]=-{\hat L}.\label{dynscalingXL}
\EEQ
Whenever, the dynamical exponent $z\ne 1$, we shall find an explicit dependence on $v$.
In general, we look for a family of generators $X_n$, for which we make the ansatz
\BEQ
X_n=-a_n(t,r,v)\partial_t-b_n(t,r,v)\partial_r-c_n(t,r,v)\partial_v-d_n(t,r,v).\label{tofind}
\EEQ
We shall find $X_n$ from the following three conditions (throughout, we use the notations $\partial_tf={\dot f}, \partial_rf=f'$):
\begin{enumerate}
\item $X_n$ must be a symmetry for the equation (\ref{ineq2}), hence $[{\hat L},X_n]=\lambda_n{\hat L}$. This gives
\BEA
&& \mu\dot{a}_n+v a'_n+\mu\lambda_n=0, \quad \mu\dot{b}_n+v b'_n-c_n+\lambda_n v=0 \label{symcond} \\
&& \mu{\dot c}_n+vc'_n=0, \quad\mu{\dot d}_n+vd'_n=0.\nonumber
\EEA
\item The generator $X_0$ is assumed to be in the Cartan sub-algebra, hence $[X_n,X_0]=\alpha_{n,0}X_n$. It follows
\BEA
(1+\alpha_{n,0})a_n-t{\dot a}_1-\frac{r}{z}a'_n-\frac{1-z}{z}v\partial_v a_n &=&0\label{line9}\\
(1/z+\alpha_{n,0})b_n-t{\dot b}_n-\frac{r}{z}b'_n-\frac{1-z}{z}v\partial_v b_n &=&0\label{line10}\\
((1-z)/z+\alpha_{n,0})c_n-t{\dot c}_1-\frac{r}{z}c'_n-\frac{1-z}{z}v\partial_v c_n &=&0\label{line11}\\
\alpha_{n,0}d_n-t{\dot d}_n-\frac{r}{z}d'_n-\frac{1-z}{z}v\partial_v d_n&=&0.\label{line12}
\EEA
\item The action of $X_{-1}$ is as a lowering operator, hence $[X_n,X_{-1}] = \alpha_{n,-1}X_{n-1}$. It follows
\BEA
{\dot a}_n &=& \alpha_{n,-1} t, \quad {\dot b}_n = \alpha_{n,-1} r/z \label{annihilation}\\
     {\dot c}_n &=& \alpha_{n,-1}v(1-z)/z, \quad {\dot d}_n = \alpha_{n,-1}x/z. \nonumber
\EEA
\end{enumerate}
These conditions, combined with the following initial conditions:
\BEA a_0 &=& t, \quad b_0=\frac{r}{z}, \quad c_0=\frac{1-z}{z}v, \quad d_0=x\nonumber\\
     a_{-1} &=& 1, \quad b_{-1}=0, \quad c_{-1}=0, \quad d_{-1}=0.\label{initialcond}
\EEA
must be sufficient for determination of all admissible forms of $X_n$.

In the special case $n=1$, we have $\alpha_{1,0}=1$ and find the most general form of $X_1$ as a symmetry of (\ref{ineq2})
as follows:\footnote{The requirement that $\langle X_{\pm 1,0}\rangle$ 
close into the Lie algebra $\mathfrak{sl}(2,\mathbb{R})$ fixes $\alpha_{1,-1}=2$.}
\BEQ
X_1 = -a_1(t,r,v)\partial_t-b_1(t,r,v)\partial_r-c_1(t,r,v)\partial_v-d_1(t,r,v) \label{x1determine}
\EEQ
and
\BEA
a_1(t,r,v) &=& t^2+A_{12}r^2v^{-2}+A_{110}rv^{\frac{2z-1}{1-z}}+A_{100}v^{\frac{2z}{1-z}}\label{a1determine}\\
b_1(t,r,v) &=& \frac{2}{z}tr+\left(\frac{A_{12}}{\mu}
+\frac{z-2}{z}\mu\right)r^2v^{-1}+B_{110}rv^{\frac{z}{1-z}}+B_{100}v^{\frac{z+1}{1-z}}\label{b1determine}\\
c_1(t,r,v) &=& \frac{2}{z}(1-z)(vt-\mu r)+(B_{110}-\frac{A_{110}}{\mu})v^{\frac{z}{1-z}}\label{c1determine}\\
d_1(t,r,v) &=& \frac{2}{z}xt -\frac{2}{z}\mu x r v^{-1}+D_0 v^{\frac{z}{1-z}}\label{d1determine}
\EEA
with a certain set of undetermined constants.

For conformal invariance, a family of generators $Y_n$ must also be found.
Its construction is straightforward if the explicit form of $Y_{-1}$ is known. Really
$X_1$ must act as a raising operator, in both hierarchies, such that \cite{Henkel02}
\BEQ
[X_1,Y_{-1}]\sim Y_0,\quad [X_1,Y_0]\sim Y_1.\label{keyym}
\EEQ
which implies that $[Y_{-1},[Y_{-1},X_1]]\sim Y_{-1}$. 
However, the usual realization of $Y_{-1}=-\partial_r$ as space translations does {\em not} work, since
if we set all undetermined constants in eq.~(\ref{x1determine}) to zero, one would have $[Y_{-1},[Y_{-1},X_1]]\sim v^{-1}Y_{-1}$.
It is better to work with the form
\BEQ
Y_{-1}=-v\partial_r.\label{yminusansatz}
\EEQ
as we shall do from now on.

\noindent We first consider the special case, when all the constants in the expression (\ref{x1determine}) for $X_1$ vanish: \\

\noindent \underline{{\bf Case A}: $A_{12}=A_{110}=A_{100}=B_{110}=B_{100}=D_0=0$}. 

\noindent {\bf Proposition 2:} {\it The six generators}
\BEA
X_{-1} & = & -\partial_t,\quad X_0= -t\partial_t -\frac{r}{z}\partial_r-\frac{1-z}{z}v\partial_v-\frac{x}{z}\nonumber\\
   X_1 & = & -t^2\partial_t-\left(\frac{2}{z}t r +\frac{z-2}{z}\mu r^2 v^{-1}\right)\partial_r-\frac{2(1-z)}{z}(vt-\mu r)\partial_v
             -\frac{2}{z}x t+\frac{2}{z}\mu x r v^{-1}\nonumber\\
Y_{-1} & = & -v\partial_r, \quad Y_0 = -(t v-\frac{\mu}{z} r)\partial_r -\frac{z-1}{z}\mu v \partial_v +\mu \frac{x}{z}\nonumber\\
   Y_1 & = & -\left(t^2v-\frac{2}{z}\mu t r -\frac{z-2}{z}\mu^2r^2v^{-1}\right)\partial_r-\frac{2}{z}(z-1)\mu(vt-\mu r)\partial_v\nonumber\\
       &   & +\frac{2}{z}\mu xt-\frac{2}{z}\mu^2xrv^{-1}\label{caseAgenconf}
\EEA
{\it span a representation of the conformal algebra (\ref{commutators}), which acts as dynamical symmetry algebra of the equation
(\ref{ineq2}), for arbitrary dynamical exponent $z$.} 

\noindent {\bf Proof:} It is readily checked that the generators (\ref{caseAgenconf}) satisfy the commutation relations (\ref{commutators}), with
$\mu\mapsto -\mu$. On the other hand, for any $f=f(t,r,v)$, one has
\BEA
&& [{\hat L}, X_{-1}] = [{\hat L}, Y_{-1}] = [{\hat L}, Y_0] = [{\hat L}, Y_1]=0\nonumber\\
&&[{\hat L}, X_0] = -{\hat L},\quad [{\hat L}, X_1] = -2t {\hat L}, \nonumber 
\EEA
which establishes the asserted dynamical symmetry. \hfill~ q.e.d.

\noindent Next, we treat the general case, when all the constants are non-zero: 

\noindent\underline{{\bf Case B}: $A_{12}\ne 0, A_{110}\ne 0, A_{100}\ne 0, B_{110}\ne 0, B_{100}\ne 0, D_0\ne 0$}.

\noindent Then the  generators are modified as follows: \newpage
\BEA  {\bar X}_1 & = & X_1 +\wit{X}_1\nonumber\\
      \wit{X}_1 & = & -\left(A_{12}r^2v^{-2}+A_{110}rv^{\frac{2z-1}{1-z}}+A_{100}v^{\frac{2z}{1-z}}\right)\partial_t\nonumber\\
                &  & -\left(\frac{A_{12}}{\mu}r^2v^{-1}+B_{110}rv^{\frac{z}{1-z}}+B_{100}v^{\frac{z+1}{1-z}}\right)\partial_r\nonumber\\
                &  & -(B_{110}-\frac{A_{110}}{\mu})v^{\frac{z}{1-z}}\partial_v- D_0 v^{\frac{z}{1-z}},\label{varx1}
\EEA
\BEA {\bar Y}_0 &=& Y_0+\wit{Y}_0\nonumber\\
      \wit{Y}_0 &=& \demi[\wit{X}_1,Y_{-1}]\nonumber\\
                &=& -(A_{12}rv^{-1}+\demi A_{110}v^{-1+1/(1-z)})\partial_t-\frac{1}{2\mu}(2A_{12}r+A_{110}v^{1/(1-z)})\partial_r.\label{vary0}
\EEA
Now, computing
\BEQ
[{\bar Y}_0, Y_{-1}]=-\mu Y_{-1}+A_{12}X_{-1}+\frac{A_{12}}{\mu}Y_{-1}\label{newcom1}
\EEQ
we conclude that the cases $A_{12}=0$ and $A_{12}\ne 0$ must be treated separately.\\

\noindent\underline{{\bf case B1}: $ A_{12}=0$}. It follows that the constants in (\ref{x1determine}) are given by:
\BEQ
B_{110}=A_{110}/\mu, \quad  A_{100} = \frac{A_{110}^2}{4\mu^2}, \quad B_{100}=\frac{A_{100}}{\mu}=\frac{A_{110}^2}{4\mu^3}, \quad D_0=0.\nonumber
\EEQ
{\bf Proposition 3:} {\it Let $z\ne 1$ and $A_{110}$ be arbitrary constants. Then the six generators}
\BEA
{\bar X}_{-1} & = & -\partial_t,\quad 
{\bar X}_0= -t\partial_t -\frac{r}{z}\partial_r-\frac{1-z}{z}v\partial_v-\frac{x}{z}\nonumber\\
{\bar X_1} & = & -(t^2+A_{110}r v^{(2z-1)/(1-z)}+\frac{A_{110}^2}{4\mu^2}v^{2z/(1-z)})\partial_t\nonumber\\
   & & -\left(\frac{2}{z}t r +\frac{z-2}{z}\mu r^2 v^{-1}+\frac{A_{110}}{\mu}r v^{z/(1-z)}
       +\frac{A_{110}^2}{4\mu^3} v^{(z+1)/(1-z)}\right)\partial_r\nonumber\\
   & & -\frac{2(1-z)}{z}(vt-\mu r)\partial_v-\frac{2}{z}x t+\frac{2}{z}\mu x r v^{-1}\nonumber\\
{\bar Y}_{-1} &=& -v\partial_r  \nonumber \\
{\bar Y}_0 &=& -\frac{A_{110}}{2}v^{z/(1-z)}\partial_t-(t v-\frac{\mu}{z} r+\frac{A_{110}}{2\mu}v^{1/(1-z)})\partial_r 
               -\frac{z-1}{z}\mu v \partial_v +\mu \frac{x}{z}\nonumber\\
{\bar Y}_1 & = & -A_{110}(tv^{z/(1-z)}-\mu r v^{(2z-1)/(1-z)})\partial_t\nonumber\\
   & & -\left(t^2v-\frac{2}{z}\mu t r -\frac{z-2}{z}\mu^2r^2v^{-1}+\frac{A_{110}}{\mu}(tv^{1/(1-z)}-\mu rv^{z/(1-z)})\right)\partial_r\nonumber\\
   & & -\frac{2}{z}(z-1)\mu(vt-\mu r)\partial_v+\frac{2}{z}\mu xt-\frac{2}{z}\mu^2xrv^{-1}\label{caseB1genconf}
\EEA
{\it span a representation of the conformal algebra.}\footnote{The above result of {\bf case A} is recovered upon setting $A_{110}=0$.}
{\it These generators give more symmetries of the equation (\ref{ineq2}).}\\

\noindent {\bf Proof:} From the above, the commutator (\ref{commutators}) are readily verified, with $\mu\mapsto -\mu$. 
For the dynamical symmetries, one checks the commutators
\BEA  [{\hat L}, X_{-1}] &=& [{\hat L}, Y_{\pm 1,0}] \:=\: 0\nonumber\\
{} [{\hat L}, X_0] &=& -{\hat L},\quad [{\hat L}, X_1] = -(2t+\frac{A_{110}}{\mu}v^{z/(1-z)}) {\hat L}. \nonumber 
\EEA
which proves the assertion. \hfill~ q.e.d. 

In contrast to the previous {\bf case A}, the representation acting only on $(t,r)$ but keeps $v$ is a constant parameter, 
can no longer be obtained by simply setting $z=1$. Rather, one must set $A_{110}=0$ first and only then the limit $z\to 1$ is well-defined. \\

\noindent\underline{{\bf case B2}: $A_{12}\ne 0, A_{110}\ne 0, B_{110}\ne 0, A_{100}\ne 0, B_{100}\ne 0, D_0\ne 0$}.

\noindent It turns out that for $A_{12}\ne 0$, the algebra also can be closed, but only if $A_{12}=\mu$ and $A_{110}=0$ 
(then all others constants also vanish).\\

\noindent {\bf Proposition 4:} {\it Let $z$ be an arbitrary constant. 
Then the generators $\left\langle {\cal X}_{\pm 1,0}, {\cal Y}_{\pm 1,0}\right\rangle$, where}
\BEA
{\cal X}_{-1} & = & -\partial_t,\quad {\cal X}_0= -t\partial_t -\frac{r}{z}\partial_r-\frac{1-z}{z}v\partial_v-\frac{x}{z}\nonumber\\
{\cal X}_{-1} &=& X_{-1},\quad {\cal X}_0=X_0\nonumber\\
{\cal X}_1 & = & -\left(t^2+\mu r^2v^{-2}\right)\partial_t- \left(\frac{2}{z}t r +\frac{z+\mu(z-2)}{z}r^2 v^{-1}\right)\partial_r\nonumber\\
   & & -\frac{2(1-z)}{z}(vt-\mu r)\partial_v-\frac{2}{z}x t+\frac{2}{z}\mu x r v^{-1}\nonumber\\
{\cal Y}_{-1} &=& -v\partial_r \nonumber \\
{\cal Y}_0 & = & -\mu r v^{-1}\partial_t-\left(t v-(\frac{\mu}{z}-1) r\right)\partial_r -\frac{z-1}{z}\mu v \partial_v +\mu \frac{x}{z}\nonumber\\
{\cal Y}_1 & = & -\mu\left(2trv^{-1}+(1-\mu)r^2v^{-2}\right)\partial_t-\left(t^2v-\frac{2}{z}(z-\mu) t r 
               +\frac{z(1-\mu)-(z-2)\mu^2}{z}r^2v^{-1}\right)\partial_r\nonumber\\
           &   & -\frac{2}{z}(z-1)\mu(vt-\mu r)\partial_v+\frac{2}{z}\mu xt-\frac{2}{z}\mu^2xrv^{-1} \label{B2genconf}
\EEA
{\it close into a Lie algebra, with the following non-zero commutation relations}
\BEA
&& [{\cal X}_n, {\cal X}_{n'}] = (n-n'){\cal X}_{n+n'}, \quad  [{\cal X}_n, {\cal Y}_m]=(n-m){\cal Y}_{n+m}\nonumber\\
&& [{\cal Y}_m, {\cal Y}_{m'}] = (m-m')\left(\mu{\cal X}_{m+m'}+(1-\mu){\cal Y}_{m+m'}\right), \label{newcomgeneral}
\EEA
{\it with $n,n',m,m'\in \{-1, 0, 1\}$.
The algebra is isomorphic to the usual conformal algebra (\ref{commutators}) and further extends the dynamical symmetries of the equation (\ref{ineq2}).}\\

\noindent {\bf Proof:} The commutation relation are directly verified. The isomorphism with the conformal algebra follows from Proposition~1.
The requirement to have an symmetry algebra of equation (\ref{ineq2}) implies a relation between
the constants $k,q$ (called $\alpha, \beta$ in Proposition~1) and $\mu$, namely $q=(k-\mu^2)/\mu$.
In this case at hand, we have $k=\mu$, $q=1-\mu $. It is then verified that $[{\hat L}, {\cal X}_{-1}] = [{\hat L}, {\cal Y}_{-1}] = 0$ and
\BEA
&&[{\hat L}, {\cal X}_0] = -{\hat L} \nonumber \\
&&[{\hat L}, {\cal X}_1] = -2(t+\frac{r}{z}v^{-1}) {\hat L}\nonumber\\
&&[{\hat L},{\cal Y}_0]= -(k/\mu){\hat L}=-{\hat L}\nonumber \\
&&[{\hat L},{\cal Y}_1]=-2\left(\frac{k}{\mu}t+\frac{k}{z\mu^2}rv^{-1}\right){\hat L}=
-2\left(t+\frac{1}{z\mu}rv^{-1}\right){\hat L}. \nonumber 
\EEA
which proves that these are dynamical symmetries of (\ref{ineq2}). \hfill~ q.e.d.

We now ask whether the finite-dimensional representations (\ref{caseAgenconf}, \ref{caseB1genconf}, \ref{B2genconf}), with $\mu\ne 0$, 
acting on functions $f=f(t,r,v)$, and having a dynamical exponent $z\ne 1$, can be extended 
to representations of an infinite-dimensional conformal Virasoro algebra. The answer turns out to be negative:

\noindent {\bf Proposition 5:}
{\it The representations (\ref{caseAgenconf}, \ref{caseB1genconf}, \ref{B2genconf}) of the finite-dimensional conformal algebra 
$\left\langle X_n,Y_n\right\rangle_{n\in\{\pm 1,0\}}$ with commutators (\ref{symmetryalgebra})
cannot be extended to representations of an infinite-dimensional conformal Virasoro algebra with commutators (\ref{symmetryalgebra}) when $z\ne 1$.}

Similar no-go results have been found before for variants of representations of the Schr\"odinger and conformal galilean algebras \cite{Henkel06b}.
On the other hand, for $\mu=0$ extensions to a representation of a conformal Virasoro algebra with $z\ne 1$ exist \cite{Cherniha04}. 
 
\noindent {\bf Proof:}
Since for the finite-dimensional representations (\ref{caseAgenconf}, \ref{caseB1genconf}, \ref{B2genconf}),  we have
\BD 
[X_n, X_{n'}]=(n-n')X_{n+n'},\quad [X_n, Y_{m}]=(n-m)Y_{n+n'}, \quad n,n',m=0,\pm 1 
\ED
we suppose that this must be valid for all admissible $n, m\in\mathbb{Z}$.
Now using the condition (\ref{annihilation}) for $n=2$, a conformal Virasoro algebra should contain a new generator $X_2$. 
Starting from the most general form, 
$X_2=-a_2(t,r,v)\partial_t-b_2(t,r,v)\partial_r-c_2(t,r,v)\partial_v-d_2(t,r,v)$ we find that the coefficients are obtained from:
\BEA
a_2 &=& t^3+a_{21}(r,v),\quad b_2=\frac{3}{z}t^2 r+3\frac{z-2}{z}\mu t r^2 v^{-1}+b_{21}(r,v)\nonumber\\
c_2 &=& 3\frac{1-z}{z}(vt^2/2-\mu r t)+c_{21}(r,v), \quad d_2=\frac{3}{z}xt^2-\frac{6}{z}\mu x +d_{21}(r,v), \nonumber 
\EEA
where $a_{21}(r,v), b_{21}(r,v), c_{21}(r,v), d_{21}(r,v)$ are unknown functions of their arguments, but do no longer depend on the time $t$.
We want to satisfy $[X_2,Y_{-1}]=3Y_1$. However when calculating
\BEA
[X_2,Y_{-1}] & = & [-a_2\partial_t-b_2\partial_r-c_2\partial_v-d_2, -v\partial_r]=\nonumber\\
             & = & 3Y_1-va'_{21}\partial_t-(3\frac{1-z}{2z}t^2v-\frac{3}{z}(1-z)\mu t r + vb'_{21}-c_{21}+3\frac{z-2}{z}\mu^2r^2v^{-1})\partial_r\nonumber\\
             &   & -(vc'_{21}+3\frac{1-z}{z}\mu(tv-2\mu r))\partial_v-vd'_{21}-\frac{6}{z}\mu\gamma r v^{-1}\nonumber
\EEA
we see that closure is not possible for $z\ne 1$. Indeed, although the dependence on $r,v$ of the functions
$a_{21}, b_{21}, c_{21}, d_{21}$ can be chosen to satisfy the above closure condition, the $t$-dependence can not be absorbed into these functions.
Hence our new representations (\ref{caseAgenconf}, \ref{caseB1genconf}, \ref{B2genconf}) of the conformal algebra
(\ref{symmetryalgebra}) are necessarily finite-dimensional. \hfill ~ q.e.d.

\section{Symmetry algebra of collisionless Boltzmann equation with an extra force term}

We write the collisionless Boltzmann equation in the form
\BEQ
{\hat B}f(t,r,v)=\left(\mu\partial_t+v\partial_r+F(t,r,v)\partial_v\right)f(t,r,v)=0.\label{eq:Boltzmann}
\EEQ
We want to determine the admissible forms of an external force $F(t,r,v)$ such that the equation (\ref{eq:Boltzmann}) is invariant under
a representation of the conformal algebra (\ref{symmetryalgebra}).
The unknown representation must include the ``force'' term and in particular for $F(t,r,v)=0$
it should coincide  with the representations of conformal algebra obtained in previous section.

The idea of the construction is similar to the one used in section~2. First, we impose invariance under basic symmetries:
\begin{itemize}
\item From invariance under time-translation $X_{-1}=-\partial_t$, it follows
\BEQ [X_{-1}, {\hat B}]=-\dot{F}=0\rightarrow F=F(r,v)\label{timeinvariance}
\EEQ
\item From invariance under dynamical scaling $X_0=-t\partial_t-\frac{r}{z}\partial_r-\frac{1-z}{z}v\partial_v-\frac{x}{z}$, we obtain that
\BEQ  [{\hat B}, X_0]=-{\hat B},\label{dynscaling}
\EEQ
if $F(r,v)$ satisfies the equation $(r\partial_r+(1-z)v\partial_v-(1-2z))F(r,v)=0$, with solution
\BEQ \label{detF}
F(r,v)=r^{1-2z} \varphi(r^{z-1}v),
\EEQ
where $\varphi(u)$ is an arbitrary function, of the scaling variable $u:=r^{z-1}v$.
\end{itemize}
It turns out that for the following calculations, it is more convenient to make a change of independent variables
$(t, r, v) \mapsto (t, r, u)$. In the new variables, the generator of dynamical scaling just reads:
\BEQ
X_0 = -t\partial_t -\frac{r}{z}\partial_r-\frac{x}{z}.\label{dynscalingX}
\EEQ
Next, in order to be specific, we make the following ansatz for the analogue of space translations\footnote{Indeed,
we might also require to find $Y_{-1}$ from the conditions to be (i) a symmetry of Boltzmann equation and
(ii) to form a closed Lie algebra with the other basic symmetries $X_{-1,0}$.
Such requirements lead to a system of differential equations and the ansatz (\ref{ansatz}) is a particular solution of this system,
which has the special property that the Boltzmann operator can be linearly expressed ${\hat B}= -\mu X_{-1}-Y_{-1}$ by the generators.
We believe this to be a natural auxiliary hypothesis.}
\BEQ \label{ansatz}
Y_{-1}= -r^{1-z}u\partial_r-r^{-z}\Phi(u)\partial_u, \quad \Phi(u)=(z-1)u^2+\varphi(u).
\EEQ
In the same coordinate system, the collisionless Boltzmann equation becomes
\BEQ
{\hat B}f(t,r,u)= \left(\mu\partial_t +r^{1-z}u\partial_r+r^{-z}\Phi(u)\partial_u\right)f(t,r,u)=0.\label{Boltzmann}
\EEQ

Here, some comments are in order. In the structure of Boltzmann equation (\ref{Boltzmann}), as well as in the form (\ref{ansatz})
of the modified space translations $Y_{-1}$, enters an unknown function $\Phi(t,r,u)$. Therefore, the form of $X_1$ cannot be found only from its commutator
with the other generators $X_n$, but the constraints form the entire conformal algebra must be used, as well as the requirement that
$X_1$ and $Y_{0,1}$ are dynamical symmetries of eq.~(\ref{Boltzmann})
\BEQ \label{newsym}
[{\hat B},X_1]=\lambda_{X_1}(t,r,v){\hat B}, \quad [{\hat B},Y_0]=\lambda_{Y_0}(t,r,v){\hat B}, \quad [{\hat B},Y_1]=\lambda_{Y_1}(t,r,v){\hat B}.
\EEQ
In fact, commuting the unknown generators $X_1, Y_0, Y_{1}$ with $X_{-1}$ and $X_0$,
we can fix the $t$- and $r$-dependence of the yet undetermined functions
which occur in them
\BEA
Y_0 &=& -r^za_0(u)\partial_t-(r^{1-z}u+rb_0(u))\partial_r-(r^{-z}\Phi(u)t+c_0(u))\partial_u-d_0(u)\nonumber\\
X_1 &=& -(t^2 +r^{2z}a_{12}(u))\partial_t - ((2/z)t r + r^{z+1}b_{12}(u))\partial_r\nonumber\\
    & & - r^zc_{12}(u)\partial_u-(2/z)xt - r^zd_{12}(u)\nonumber\\
Y_1 & = & -\left(2t r^za_0(u)+r^{2z}A(u)\right)\partial_t-\left(t^2r^{1-z}u+2trb_0(u)+r^{z+1}B(u)\right)\partial_r\nonumber\\
    & & -\left(t^2r^{-z}\Phi(u)+2tc_0(u)+r^zC(u)\right)\partial_u+
         (2/z)\mu x t - r^zD(u),\label{newgenconf}
\EEA
with the four functions
\BEA
A(u) & = & 2zb_0a_{12}+c_0a'_{12}-za_0b_{12}-a'_0c_{12}, \quad C(u) = zb_0c_{12}+c_0c'_{12}-c'_0c_{12}-a_{12}\Phi\nonumber\\
B(u) & = & \frac{2}{z}a_0+zb_0b_{12}+c_0b'_{12}-ua'_{12}-b'_0c_{12}, \quad D(u) = \frac{2}{z}x a_0+zb_0d_{12}+c_0d'_{12}.\label{relationx1y0y1}
\EEA
In particular, looking for a representation of the analog of extended Galilei algebra $\langle X_{-1}, X_0, Y_{-1}, Y_0\rangle$,
we find that the unknown functions
$a_0(u), b_0(u), c_0(u), d_0(u)$ must satisfy the system
\BEA  z u a_0(u)+\Phi(u) a'_0(u)-k &=&0\label{eq1}\\
       z u b_0(u)+\Phi(u) b'_0(u)-c_0(u)-qu &=&0\label{eq2}\\
       \Phi'(u)c_0-\Phi(u)c'_0(u)+(q-zb_0)\Phi&=&0\label{eq3}\\
       \Phi(u)d'_0(u)&=&0 \label{eq3bis}
\EEA
Because of (\ref{eq3bis}), one must distinguish two cases:
\begin{enumerate}
\item $\Phi(u)=0$, when $d_0(u)$ can be arbitrary
\item $\Phi(u)\ne 0$, when $d_0(u)=d_0=\mbox{\rm cste.}$ is a constant.
\end{enumerate}
In the second case, taking equation (\ref{eq2}, \ref{eq3}) together, we obtain an equation for $b_0(u)$. It is
\BEQ
\Phi^2(u)b''_0(u)+zu\Phi(u)b'_0(u)+(2z\Phi(u)-zu\Phi'(u))b_0(u)-2s\Phi(u)=0, \label{detb0}
\EEQ
and has in general two independent solution: $b_{01}(u),b_{02}(u)$. It follows that,
for a given arbitrary value of $\Phi(u)\ne 0$, we have in general {\em two distinct realisations}
of the analogue of Galilei transformation;
and consequently also two realizations of the analogue of the Galilei algebra. By construction, these
are Lie algebras of  symmetries of the collisionless Boltzmann equation (\ref{Boltzmann}) (with $\lambda_{Y_0}=-k/\mu=-(\mu +q)$):
\BEA
  && [Y_0, X_{-1}]=Y_{-1}, \quad [X_0,X_{-1}]=X_{-1}, \quad \nonumber\\
  && [Y_0, Y_{-1}]= \frac{k-\mu^2}{\mu}Y_{-1}+kX_{-1}.\label{relativistgalilei}
\EEA
Next, we include the generators of special conformal transformation $X_1$ and $Y_1$
to the extended Galilei algebras (\ref{relativistgalilei}) just constructed.
We must also satisfy the other commutators of the conformal algebra (\ref{symmetryalgebra}).
Furthermore, the generators of the representation
we are going to construct are dynamical symmetries of the collisionless Boltzmann equation.\footnote{We use the commutators $[Y_1,Y_0]=KX_1+QY_1$ and
$[Y_1,Y_{-1}]=k_0X_0+q_0Y_0$ in order to establish a relation between the constants $k,q$ and $K,Q,k_0,q_0$.}
We find:
\BEQ
\lambda_{X_1}(t,r,u) = -2t-(r^z/\mu)(2zua_{12}+\Phi(u)a'_{12}(u))=-2t -2r^za_0(u)/\mu
\EEQ
for the eigenvalue and
\BEA
     && c_{12}(u)=(2/z)\mu - (u/\mu)(2za_{12}(u)+\Phi(u)a'_{12}(u))+(2zub_{12}+\Phi(u)b'_{12}(u))\label{eq4}\\
     && zuc_{12}(u)+\Phi c'_{12}(u)-c_{12}(u)\Phi'(u)+zb_{12}(u)\Phi(u)-2c_0(u)=0\label{eq5}\\
     && zud_{12}(u)+\Phi(u)d'_{12}(u)+(2/z)\mu x = 0\label{eq6}\\
     && \Phi^2(u)b''_{12}(u)+3zu\Phi(u)b'_{12}(u)+z[2zu^2+3\Phi(u)-2u\Phi'(u)]b_{12}(u)\nonumber\\
     && ~~-(u/\mu)\Phi^2(u)a''_{12}(u)-[3zu^2+2\Phi(u)](\Phi/\mu)a'_{12}(u)\nonumber\\
     && ~~-[zu^2+3\Phi(u)-u\Phi'(u)](2zu/\mu)a_{12}(u)+(2\mu/z)(z u-\Phi'(u))=0\label{eq7}\\
     && 2zua_{12}(u)+\Phi(u)a'_{12}(u)-2a_0(u)=0\label{eq8}\\
     && 2zub_{12}(u)+\Phi(u)b'_{12}(u)-c_{12}(u)-2b_0(u)=0\label{eq9}\\
     && b_0(u)=(u/\mu)a_0(u)-\mu/z\label{eq10}\\
     && c_0(u)=(\Phi/\mu)a_0(u)\label{eq11}\\
     && d_0(u)=\mbox{\rm cste.}=-\mu x/z.\label{eq12}\\
     && k_0 = \alpha_0k=2k, \quad q_0 = \alpha_0q=2q\nonumber\\
     && 2z u A(u)+\Phi(u)A'(u)-2qa_0(u)=0\label{eq13}\\
     && 2z u B(u)+\Phi(u)B'(u)-C(u)-2(k/z+qb_0(u))=0\label{eq14}\\
     && z u C(u)+\Phi(u)C'(u)-\Phi'(u)C(u)+z\Phi(u)B(u)-2qc_0(u)=0\label{eq15}\\
     && z u D(u)+\Phi(u)D'(u)-(2x/z)(k-\mu q)=0.\label{eq16}\\
     && K=k, \quad Q=q \label{eq16bis}\\
     &&(q-2zb_0)A(u)-c_0A'(u)+za_0(u)B(u)+a'_0(u)C(u)+ka_{12}(u)-2a^2_0=0\label{eq17}\\
     &&(q-zb_0)B(u)-c_0B'(u)+u A(u)+b'_0C(u)+kb_{12}(u)-2a_0(u)b_0(u)=0\label{eq18}\\
     &&(q-zb_0+c'_0(u))C(u)-c_0C'(u)+\Phi(u) A(u)+kc_{12}(u)-2a_0(u)c_0(u)=0\label{eq19}\\
     &&(q-zb_0)D(u)-c_0D'(u)+kd_{12}(u)+\frac{2a_0(u)\mu x}{z}=0\label{eq20}\\
     && 2z(b_{12}(u)A(u)-a_{12}(u)B(u))+c_{12}(u)A'(u)-a'_{12}(u)C(u)+2a_0(u)a_{12}(u)=0
    \label{eq21}\\
          && (2/z)A(u)-c_{12}(u)B'(u)+b'_{12}C(u)-2b_0(u)a_{12}(u)=0\label{eq22}\\
          && (zb_{12}(u)-c'_{12}(u))C(u)+c_{12}C'(u)-zc_{12}(u)B(u)+2c_0(u)a_{12}(u)=0
          \label{eq23}\\
          && (2x/z)( \mu a_{12}(u)+A(u))+zd_{12}(u)B(u)+d'_{12}(u)C(u)\nonumber\\
          && ~~-zb_{12}D(u)-c_{12}(u)D'(u)=0\label{eq24}
    \EEA
The system of equations
(\ref{eq1}, \ref{eq2}, \ref{eq3}, \ref{eq4}, \ref{eq5}, \ref{eq6}, \ref{eq7}, \ref{eq8}, \ref{eq9}, \ref{eq10}, \ref{eq11},
\ref{eq12}, \ref{eq13}, \ref{eq14}, \ref{eq15}, \ref{eq16}, \ref{eq16bis}, \ref{eq17}, \ref{eq18}, \ref{eq19}, \ref{eq20}, \ref{eq21}, \ref{eq22},
\ref{eq23}, \ref{eq24}) must give a solution for the unknown functions $a_0(u),b_0(u),c_0(u),d_0(u),a_{12}(u),b_{12}(u),c_{12}(u),d_{12}(u)$.
Of course, it is possible that several of the above equations are equivalent. 
Because of this fact, although the above system might look to be over-determined,
we have not yet been able to produce explicit solution without making auxiliary assumption.
A classification of all solutions of the above system is left as an open problem.
We shall now describe some examples of solutions of this large system.

\noindent\underline{{\bf Example 1:} Let $\Phi(u)=0$}.
This case seems to be quite simple, provided it is compatible with our system. From equation
(\ref{eq1}) we obtain
\BEQ
a_0(u)=\frac{k}{z}u^{-1} \label{sol1}
\EEQ
Using this value of $a_0(u)$ from equations (\ref{eq6}, \ref{eq7}, \ref{eq8}, \ref{eq10}, \ref{eq11}, \ref{eq12}), we directly obtain
\BEA
 b_0 & = & \mbox{\rm cste.}=\frac{k}{z\mu}-\frac{\mu}{z}, \quad c_0(u) = 0, \quad d_0 = \mbox{\rm cste.} = -\frac{\mu}{z}x,\label{soly0}\\
a_{12}(u) & = & \frac{k}{z^2}u^{-2}, \quad b_{12}(u) = \frac{1}{\mu z^2}(k-\mu^2)u^{-1}, \quad c_{12}(u) = 0, \quad
d_{12}(u)= -\frac{2\mu x}{z^2}u^{-1}.~~ \label{solx1}
\EEA
When we substitute the above results in relations (\ref{relationx1y0y1}), we also find
\BEA
 A(u) & = & \frac{k}{\mu z^2}(k-\mu^2)u^{-2}, \quad B(u) = \frac{1}{\mu^2 z^2}\left(k(k-\mu^2)+\mu^4\right)u^{-1}\nonumber\\
 C(u) & = & 0, \quad D(u) = \frac{2\mu^2x}{z^2}u^{-1}.\label{soly1}
\EEA
One can now verify that the above results for the functions $a_0(u),b_0(u),c_0(u),d_0(u)$ and $a_{12}(u)$, $b_{12}(u)$,$c_{12}(u),d_{12}(u),A(u),B(u),C(u),D(u)$
satisfy all equations of the above system. Now, we can finally write the algebra generators
\BEA
X_{-1} & = & -\partial_t,\quad X_0 = -t\partial_t-\frac{r}{z}\partial_r-\frac{x}{z}\nonumber\\
X_1 & = & -\left(t^2+\frac{k}{z^2}r^{2z}u^{-2}\right)\partial_t - \left(\frac{2}{z}t r +\frac{k-\mu^2}{z^2\mu}r^{z+1}u^{-1}\right)\partial_r
-\frac{2}{z}x t + \frac{2\mu x}{z^2}r^z u^{-1},\nonumber\\
Y_{-1} & = & - r^{1-z}u\partial_r,\nonumber\\
Y_{0} & = & -\frac{k}{z}r^z u^{-1}\partial_t-\left(t r^{1-z} u + \frac{k-\mu^2}{z\mu }r\right)\partial_r +\frac{\mu x}{z},\nonumber\\
Y_{1} & = & -\left(\frac{2k}{z}t r^z u^{-1}+\frac{k(k-\mu^2)}{z^2\mu}r^{2z}u^{-2}\right)\partial_t
\label{tvugenerators}\\
 && - \left(t^2r^{1-z}u+2\frac{k-\mu^2}{z\mu}t r +\frac{k(k-\mu^2)+\mu^4}{z^2\mu^2}r^{z+1}u^{-1}\right)\partial_r
 + \frac{2}{z}\mu x t - \frac{2\mu^2x}{z^2}r^z u^{-1}. \nonumber
\EEA
Returning to the original variables via the change $(t,r,u)\mapsto (t,r,v)$, done through the substitutions $u\to r^{z-1}v$ and
$\partial_r\to \partial_r+(1-z)r^{-1}v\partial_v$.
Finally we have the following representation of a conformal symmetry algebra of the collisionless Boltzmann equation (\ref{eq:Boltzmann}):
\newpage
\BEA
X_{-1} & = & -\partial_t,\quad X_0 = -t\partial_t-\frac{r}{z}\partial_r-\frac{1-z}{z}v\partial_v-\frac{x}{z}\nonumber\\
X_1 & = & -\left(t^2+\frac{k}{z^2}r^2v^{-2}\right)\partial_t - \left(\frac{2}{z}t r +\frac{k-\mu^2}{z^2\mu}r^2v^{-1}\right)\partial_r\nonumber\\
    & - & (1-z)\left(\frac{2}{z}t v +\frac{k-\mu^2}{z^2\mu}r\right)\partial_v -\frac{2}{z}x t + \frac{2\mu x}{z^2}r v^{-1},\nonumber\\
Y_{-1} & = & - v\partial_r - (1-z)r^{-1} v^2\partial_v,\nonumber\\
Y_{0} & = & -\frac{k}{z}r v^{-1}\partial_t-\left(t v + \frac{k-\mu^2}{z\mu }r\right)\partial_r -(1-z)\left(t r^{-1}v^2 
            + \frac{k-\mu^2}{z\mu }v\right)\partial_v +\frac{\mu x}{z},\nonumber\\
Y_{1} & = & -\left(\frac{2k}{z}t r v^{-1}+\frac{k(k-\mu^2)}{z^2\mu}r^2v^{-2}\right)\partial_t 
            - \left(t^2v+2\frac{k-\mu^2}{z\mu}t r + \frac{k(k-\mu^2)+\mu^4}{z^2\mu^2}r^2v^{-1}\right)\partial_r\nonumber\\
      &  & -(1-z)\left(t^2r^{-1}v^2+2\frac{k-\mu^2}{z\mu}t v + \frac{k(k-\mu^2)+\mu^4}{z^2\mu^2}r\right)\partial_v
      + \frac{2}{z}\mu x t - \frac{2\mu^2x}{z^2}r v^{-1}.\label{Cgenconf}
\EEA
{\bf Proposition 6:}
{\it The generators (\ref{Cgenconf}) close into the following Lie algebra}
\BEA && [X_n, X_{n'}]=(n-n')X_{n+n'}, \quad [X_n, Y_m] = (n-m)Y_{n+m}\nonumber\\
     && [Y_m, Y_{m'}] = (m-m')\left(k X_{m+m'} + \frac{k-\mu^2}{\mu}Y_{m+m'}\right),
\label{Cfinitconf}
\EEA 
{\it for $n, n', m, m'$ $\in $ $\{-1, 0, 1\}$ and for an arbitrary dynamical exponent $z$. They give a representation
of the finite-dimensional conformal algebra, which acts as dynamical symmetry algebra of the Boltzmann equation in the form:}
\BEQ
{\hat B}f(t, r, v)=(\mu\partial_t+v\partial_r+(1-z)r^{-1}v^2\partial_v)f(t, r, v) = 0.\label{CBoltzmann}
\EEQ

\noindent {\bf Proof:} 
The commutation relations (\ref{Cfinitconf}) are directly checked. From the commutators $[{\hat B}, X_{-1}] = [{\hat B}, Y_{-1}]=0$ and
\BEA
&&[{\hat B}, X_0] = -{\hat B},\quad [{\hat B}, X_1] = -2\left(t+\frac{k}{z\mu}rv^{-1}\right) {\hat B}\nonumber\\
&& [{\hat B},Y_0]= -(k/\mu){\hat B}, \quad [{\hat B},Y_1]=-2\left(\frac{k}{\mu}t+\frac{k}{z\mu^2}rv^{-1}\right){\hat B}.
\nonumber 
\EEA
it is seen that they generate dynamical symmetries. \hfill~ q.e.d. 

\noindent\underline{{\bf Example 2}: Let $k=0$}. In this case, $\Phi(u)$ left arbitrary, which leads to $a_0=0$ from (\ref{eq1}) and
\BEQ b_0=\mbox{\rm cste.} =-\mu/z, \quad c_0=0, \quad d_0=-\mu x/z\label{Dy0}
\EEQ
Then from eq.~(\ref{relationx1y0y1}), we obtain
\BEA
 A(u) & = & -2\mu a_{12}(u), \quad B(u)= -\mu b_{12}(u)-ua'_{12}(u)\nonumber\\
 C(u) & = & -\mu c_{12}(u)-a_{12}(u)\Phi(u), \quad D(u) = -\mu d_{12}.\label{Dy1}
\EEA
However, when substituting in eq.(\ref{eq17}, \ref{eq18}, \ref{eq19}, \ref{eq20}), taking also into account that $q=-\mu$, we find that $A(u)=a_{12}(u)=0$.
Then it is easy to check that the conditions (\ref{eq21}, \ref{eq22}, \ref{eq23}, \ref{eq24}) are fulfilled. This allows us to formulate\\

\noindent {\bf Proposition 7:} {\it Let $\Phi(u)=(z-1)u^2+\varphi(u)$. Consider the generators}
\BEA
X_{-1} & = & -\partial_t,\quad X_0 = -t\partial_t-\frac{r}{z}\partial_r-\frac{1-z}{z}v\partial_v-\frac{x}{z}\nonumber\\
X_1 & = & -t^2\partial_t - \left(\frac{2}{z}t r +r^{z+1}b_{12}(u)\right)\partial_r\nonumber\\
    & - & (1-z)\left(\frac{2}{z}t v +r^z vb_{12}(u)+\frac{r^{1-2z}}{1-z}c_{12}(u)\right)\partial_v -\frac{2}{z}x t -r^zd_{12}(u),\nonumber\\
Y_{-1} & = & - v\partial_r - (1-z)\left(r^{-1} v^2+\frac{r^{1-2z}}{1-z}\Phi(u)\right)\partial_v = - v\partial_r-r^{1-2z}\varphi(u)\partial_v,\nonumber\\
Y_{0} & = & -\left(tv - \frac{\mu}{z}r\right)\partial_r -(1-z)\left(\frac{r^{1-2z}}{1-z}\varphi(u)t-\frac{\mu}{z}v\right)\partial_v
+\frac{\mu x}{z},\nonumber\\
Y_{1} & = & - \left(t^2v-2\frac{\mu}{z}t r -\mu r^{z+1}b_{12}(u)\right)\partial_r + \frac{2}{z}\mu x t +\mu r^zd_{12}(u)
\label{Dgenconf} \\
      &  & -(1-z)\left(t^2\frac{r^{1-2z}}{1-z}\varphi(u)-\frac{2}{z}\mu tv-n\mu r^z v b_{12}(u)-\mu\frac{r^{1-2z}}{1-z}c_{12}(u)\right)\partial_v,
      \nonumber
\EEA
{\it where $c_{12}(u)=2zub_{12}(u)+((z-1)u^2+\varphi(u))b'_{12}(u)+2\mu/z$ and $\varphi(u), b_{12}(u), d_{12}(u)$ satisfy}
\BEA
[(z-1)u^2+\varphi(u)]^2 b''_{12}(u)+3zu[(z-1)u^2+\varphi(u)]b'_{12}(u) & & \nonumber\\
 +z[(z+1)u^2-2u\varphi'(u)+3\varphi(u)]b_{12}(u)+[(2-z)u-\varphi'(u)]2\mu/z&=&0\label{b12red}\\
     zud_{12}(u)+[(z-1)u^2+\varphi(u)]d'_{12}(u)+2\mu x/z&=&0.\label{d12red}
\EEA
{\it For any triplet $(\varphi(u), b_{12}(u), d_{12}(u))$ which gives a solution of the system (\ref{b12red}, \ref{d12red}),
the generators (\ref{Dgenconf}) close into the following Lie algebra}
\BEA && [X_n, X_{n'}]=(n-n')X_{n+n'}, \quad [X_n, Y_m] = (n-m)Y_{n+m}\nonumber\\
     && [Y_m, Y_{m'}] = -\mu\,(m-m') Y_{m+m'},
\label{finip0tvconf}
\EEA 
{\it for $n, n', m, m'$ $\in $ $\{-1, 0, 1\}$ and for an arbitrary constant $z$. Eq.~(\ref{Dgenconf}) is a representation
of the finite-dimensional conformal algebra and acts as dynamical symmetry algebra of the Vlasov-Boltzmann equation, with a quite general ``force'' term:}
\BEQ
{\hat B}f(t, r, v)=(\mu\partial_t+v\partial_r+r^{1-2z}\varphi(u)\partial_v)f(t, r, v) = 0.\nonumber
\EEQ
{\bf Proof:}
The commutators are satisfied for $k=0$ and $q=-\mu$ if conditions (\ref{b12red},\ref{d12red}) are fulfilled.
Under the same conditions, the symmetries are proven by the relations
\BEA
&& [{\hat B}, X_{-1}] = [{\hat B}, Y_{-1}] = [{\hat B},Y_0] = [{\hat B},Y_1]= 0\nonumber\\
&&[{\hat B}, X_0] = -{\hat B},\quad [{\hat B}, X_1] = -2t{\hat B}.\nonumber 
\EEA
~\hfill~ q.e.d. 

In particular, if we implement the physical requirement that the ``force'' term should depend only on the positions $r$,
that is $\varphi(u)=\varphi_0=\mbox{\rm cste.}$, we can compute explicitly the representation of the algebra (\ref{Dgenconf}). To do this, one must
find a solution of the system
\BEA
      [(z-1)u^2+\varphi_0]^2 b''_{12}(u)+3zu[(z-1)u^2+\varphi_0]b'_{12}(u) &&\nonumber\\
      +z[(z+1)u^2+3\varphi_0]b_{12}(u)+2\mu\frac{2-z}{z}u&=&0\label{physb12red}\\
      zud_{12}(u)+[(z-1)u^2+\varphi_0]d'_{12}(u)+2\mu x/z&=&0.\label{findd12}
\EEA
The solution of the second equation is relatively simple, even for an arbitrary $z$
\BEQ
d_{12}(u)=-\delta_0[(z-1)u^2+\varphi_0]^{\frac{z}{2(1-z)}}\int_{\R} \!\D u\: [(z-1)u^2+\varphi_0]^{\frac{z-2}{2(1-z)}}, \quad
\delta_0=\mbox{\rm cste.}\label{d12z}
\EEQ
The solution of the equation (\ref{physb12red}) for an arbitrary $z$ can be expressed in terms of hypergeometric functions, but we shall not give
its explicit form here. However, for $z=2$, the system (\ref{physb12red}, \ref{findd12}) has an elementary solution
\BEA
b_{12}(u) &=& b_{120}\frac{u}{(u^2+\varphi_0)^2}+b_{121}\frac{u^2-\varphi_0}{(u^2+\varphi_0)^2}, \quad
b_{120}=\mbox{\rm cste.}, b_{121}=\mbox{\rm cste.}\nonumber\\
d_{12}(u) &=& -\mu x \frac{u}{u^2+\varphi_0}.\label{z2b12d12}
\EEA
Substituting this into the generators (\ref{Dgenconf}) for $z=2$, gives a finite-dimensional representation of the dynamical conformal symmetry
of a collisionless Boltzmann equation of the form
\BEQ
{\hat B}f(t, r, v) = (\mu\partial_t+v\partial_r+\varphi_0r^{-3}\partial_v)f(t, r, v) = 0.\label{Dz2Boltzmann}
\EEQ

\section{Conclusions}
In this work, we have described the results of a first exploration of dynamical symmetries of collisionless Vlasov-Boltzmann transport equations.
Our main finding is that these equations admit conformal dynamical symmetries, although it does not seem to be possible to extend this to
infinite-dimensional conformal Virasoro symmetries (not even in the case of $d=1$ space dimensions). These conformal symmetries are new representations
of the conformal algebra and are inequivalent to the standard representation which is habitually used in conformal field-theory descriptions of
equilibrium critical phenomena.
Our first class of new symmetries was found by admitting the momentum $p$ (or equivalently the velocity $v=p/\mu$) as an additional independent variable,
leading to the representations (\ref{caseAgenconf}, \ref{caseB1genconf}, \ref{B2genconf}).
The second class of symmetries also allowed for external driving forces $F(t,r,v)$ and it has been one of the questions which types of forces
should be compatible with conformal invariance. As an example,
we have seen that time-independent forces $F(r,v)=r^{1-2z}\vph(r^{z-1}v)$, with an arbitrary scaling function $\vph$, are admissible, and lead
to the general representation (\ref{Dgenconf}). However, the solutions of the associated system of equations for the coefficients have not yet been
classified and the complete content of these representations remains to be worked out in the future.

Some intuition can be gleaned from some examples. We have written down the explicit representations for the force $F(r,v)=(1-z)r^{-1}v^2$,
with an $z>1$ arbitrary  (\ref{Cgenconf}), and for $F(r,v)=\varphi_0\,r^{1-2z}$ (\ref{Dgenconf}, \ref{physb12red}, \ref{findd12}) with an arbitrary $z>1$.
In the later case, which could be related to physical situations, we have given the explicit representation of conformal algebra for $z=2$, when
$F(r,v)=\varphi_0r^{-3}$ (\ref{Dgenconf}, \ref{z2b12d12}). 
Having identified these symmetries, the next step would be to use these to find either exact solutions \cite{Fush93}
or else to use the algebra representations for fixing the form of co-variant $n$-point correlation functions, 
in analogy with time-dependent critical phenomena, see e.g. \cite{Henkel10}.

The results derived here can be used as a starting point to derive forms of the transition rates $w$ in the collision terms which would be compatible
with the dynamical symmetries of the collision-free equations. This kind of approach would be analogous to the one used for finding dynamical
symmetries of non-linear Schr\"odinger equations, see e.g. \cite{Boyer76,Stoimenov05}. We hope to return to this elsewhere.

\noindent {\bf Acknowledgments}
Most of the work on this paper was done during the visits of S.S. at
Universit\'e de Lorraine Nancy and of M.H. at XI$^{\rm th}$ International workshop
{\it ``Lie theories and its applications in physics''}.
These visits was supported by PHC Rila.
M.H. was partly supported by the Coll\`ege Doctoral Nancy-Leipzig-Coventry
(Syst\`emes complexes \`a l'\'equilibre et hors \'equilibre) of UFA-DFH.

\end{document}